\documentclass[%
 aip,
 amsmath,amssymb,
 reprint,%
]{revtex4-1}

\usepackage{graphicx}
\usepackage{dcolumn}
\usepackage{bm}

\usepackage[utf8]{inputenc}
\usepackage[T1]{fontenc}
\usepackage{mathptmx}
\usepackage{etoolbox}
\usepackage{float}
\usepackage{verbatim}
\usepackage{CJK}
\usepackage{indentfirst}
\usepackage{amsmath}
\usepackage{amsthm}
\usepackage{graphicx}
\usepackage{subfigure}
\usepackage{tabularx}
\usepackage{multirow}
\usepackage{algorithm}
\usepackage{algpseudocode}
\usepackage{xcolor}

\makeatletter
\def\@email#1#2{%
 \endgroup
 \patchcmd{\titleblock@produce}
  {\frontmatter@RRAPformat}
  {\frontmatter@RRAPformat{\produce@RRAP{*#1\href{mailto:#2}{#2}}}\frontmatter@RRAPformat}
  {}{}
}%
\makeatother
\begin{document}

\preprint{AIP/123-QED}

\title{Evolution of cooperation in enterprise resource-sharing networks with dynamic participation and adaptive decision-making}

\author{Pengfei Liu}
 \affiliation{
 School of Computer Science and Engineering, University of Electronic Science and Technology of China, Chengdu 611731, China.
 }

\author{Bin Pi}
 \affiliation{
 School of Mathematical Sciences, University of Electronic Science and Technology of China, Chengdu 611731, China.
 }

\author{Qin Li}
 \email{qinli1022@swu.edu.cn}
 \altaffiliation{Corresponding author}
 \affiliation{
 Business College, Southwest University, Chongqing, 402460, China.
 }
 
\author{Minyu Feng}
\affiliation{ 
College of Artificial Intelligence, Southwest University, Chongqing 400715, China.
}

 \author{Attila Szolnoki}
 \affiliation{
 Institute of Technical Physics and Materials Science, Centre for Energy Research, P.O. Box 49, H-1525 Budapest, Hungary.
 }


\date{\today}

\begin{abstract}

Resource sharing among enterprises is an important mechanism for improving resource utilization efficiency and facilitating collaborative development. However, the emergence and sustainability of cooperative behavior depend critically on whether enterprises choose to participate in resource-sharing activities and how they balance cooperation and competition in their decision-making processes. To investigate these issues, we develop an evolutionary cooperation model on the enterprise resource-sharing network that integrates participation status, decision-making modes, and cooperative strategies within a unified framework. Specifically, enterprises dynamically switch between active and dormant states according to their aspiration levels, while Q-learning is employed to enable adaptive selection between payoff-driven and conformist decision-making modes. Meanwhile, interactions between resource-sharing and non-sharing in active enterprises are characterized by the snowdrift game. Through extensive simulations, we systematically examine the effects of the payoff parameter, the reinforcement-learning parameters, the participation and Q-learning mechanism, and network size on enterprise participation and the evolution of cooperation. The results show that an intermediate aspiration level is most favorable for sustaining enterprise participation and facilitating the emergence of cooperation. Moreover, the dynamic activation mechanism effectively mitigates the decline in cooperation under high cost-to-benefit ratios, allowing the system to sustain a decent level of cooperative behavior. These findings highlight the important roles of dynamic participation and adaptive decision-making in shaping the evolution of resource-sharing cooperation and provide a new modeling framework to understand the emergence and persistence of cooperation in complex enterprise networks.

\end{abstract}

\maketitle

\begin{quotation}

Resource sharing represents an important form of cooperative behavior in complex systems. By exchanging resources, information, and capabilities, individuals and organizations can establish interdependencies, improve resource utilization, and generate mutual benefits. In real-world enterprise networks, however, enterprises 
can
dynamically enter or withdraw from interactions according to their expected benefits and 
can
adopt different decision-making mechanisms. To capture these features, we develop an evolutionary game model that integrates dynamic participation with reinforcement learning-based decision-making and investigate their coupled effects on the evolution of cooperation. Our results show that dynamic participation can reshape the evolution of cooperation and mitigate its decline under unfavorable payoff conditions. These findings demonstrate the importance of incorporating both endogenous participation dynamics and adaptive decision-making into the study of cooperation in complex networks.

\end{quotation}

\section{Introduction}\label{sec:introduction}

As interconnections and collaboration among enterprises continue to deepen, resource sharing has become an important means of improving resource utilization efficiency, reducing operating costs, and promoting inter-enterprise synergy~\cite{huo2025impact, xu2024impact}. In a resource-sharing network, enterprises establish local interactions through social ties or resource-exchange relationships. By sharing idle or available resources, such as production capacity, energy, information, and computing resources, enterprises can overcome individual resource constraints and facilitate the cross-enterprise allocation and utilization of resources~\cite{stojcic2026resource, li2021promoting}. Resource sharing can therefore generate not only direct economic benefits for participating enterprises but also broader network-level benefits through cooperation. However, such cooperative relationships are not inherently stable. On the one hand, resource sharing usually involves costs, requiring enterprises to invest resources, time, or capital to support others~\cite{yadav2025circular}. On the other hand, enterprises 
can
benefit from the resources shared by others without contributing themselves, 
creating
a classic free-rider problem~\cite{feldman2005overcoming, niu2025addressing}. This tension between individual and collective interests makes 
this
resource sharing a typical social dilemma. Understanding how resource-sharing cooperation emerges, evolves, and persists is therefore important 
to reveal
the mechanisms underlying cooperation in complex enterprise networks.

Evolutionary game theory provides 
an adequate
framework for studying such social dilemmas and has been widely used to investigate the emergence and evolution of cooperation~\cite{sigmund1999evolutionary, han2024selection, li2020evolution}. In a resource-sharing network, enterprises 
may either share or retain their resources.
Enterprises that share resources bear certain costs while generating benefits for their interaction partners, whereas non-sharing enterprises may still benefit from the sharing behavior of others. This payoff structure is well captured by the snowdrift game~\cite{hauert2004spatial, li2026memory, xu2023higher}. Previous studies of evolutionary games on networks have demonstrated that network topology~\cite{wang2026public, roca2009effect, zeng2025bursty}, local interactions~\cite{han2025public, zeng2025evolutionary}, and payoff structures~\cite{basak2024evolution, pi2026evolutionary} can strongly affect the evolution of cooperation. However, most existing models assume that individuals participate continuously in network interactions and that their decision-making mechanisms remain fixed throughout the evolutionary process. These assumptions cannot adequately reflect real-world enterprise resource-sharing systems, in which enterprises can dynamically adjust their participation according to their operating conditions and external environment, while different enterprises may also adopt different decision-making modes under the same resource-sharing conditions~\cite{ma2026technology, zhao2025decision}.

Beyond the evolution of cooperation and defection strategies, participation in social interactions can itself be an important behavioral component of evolutionary games. In models of voluntary participation, individuals can abstain from interactions rather than continuously engage in the underlying social dilemma. For example, Cardinot~{\it et al.} introduced probabilistic abstention and showed that intermediate levels of abstention can promote cooperation under adverse conditions~\cite{cardinot2018cooperation}. More generally, Perc and Szolnoki reviewed coevolutionary games in which strategies evolve alongside additional individual or interaction-related properties, emphasizing the importance of considering multiple coupled evolutionary dimensions when investigating cooperation~\cite{perc2010coevolutionary}. Motivated by this line of research, we incorporate dynamic participation into a resource-sharing game among enterprises. Enterprises can participate in or withdraw from resource-sharing activities depending on factors such as realized benefits, participation costs, and satisfaction with the sharing environment~\cite{li2024research, liu2025does, feng2024promote}. If an enterprise receives insufficient returns over an extended period, its willingness to participate may decline, eventually leading to withdrawing from resource-sharing interactions. Conversely, changes in the participation states and strategies of neighboring enterprises may create favorable conditions for previously dormant enterprises to re-enter the system. In this study, enterprises participating in the resource-sharing game are defined as active, whereas those temporarily not participating are defined as dormant. Changes in participation status modify the set of effective interaction partners, which in turn affects payoff accumulation and the evolution of cooperation. This creates a feedback mechanism between participation dynamics and cooperative behavior.

Differences among enterprises are reflected not only in the resource-sharing strategies they adopt but also in the way these strategies are selected. Enterprises in resource-sharing networks can adopt different decision-making rules when deciding whether to share resources. On the one hand, they can prioritize individual economic interests and imitate strategies associated with higher payoffs~\cite{kastelic2024cooperation, gimenez2011entrepreneurial}. On the other hand, they can adapt to local social norms by aligning their strategies with those of neighboring enterprises~\cite{zheng2025once, zeng2026conformist}. To capture these two complementary mechanisms, we consider payoff-driven and conformist decision-making modes. Importantly, these modes are not assumed to be fixed attributes of enterprises, but rather adaptive behavioral patterns that can change in response to local conditions.

Reinforcement learning provides a suitable framework for describing such adaptive decision-making~\cite{xie_k_amc26, shen_y_csf24, zhang_hy_c26, feng2026ai}. In particular, Q-learning enables enterprises to evaluate the long-term consequences of different decision-making modes through repeated interactions with the environment, without requiring behavioral preferences to be prescribed in advance~\cite{clifton2020q, pi2025dynamic, yan2024periodic}. Recent studies have incorporated Q-learning into evolutionary games to investigate how local interaction information, state transitions, and network structure influence the evolution of cooperation~\cite{yang2024interaction, yuan2026dynamics}. In contrast to those approaches in which Q-learning directly governs cooperation or defection, our model employs Q-learning to adaptively select between two decision-making modes. It is important to distinguish between resource-sharing strategies and decision-making modes, which represent two different levels of enterprise behavior. The former determines whether an enterprise shares resources, whereas the latter determines how that strategic choice is updated, i.e., whether the enterprise relies primarily on payoff comparisons or on neighboring behavior.

Based on these considerations, we construct an evolutionary game model for enterprise resource-sharing networks that integrates dynamic participation, adaptive decision-making, and resource-sharing strategy evolution within a unified framework. Each node represents an enterprise, and network links represent potential resource-sharing relationships. Active enterprises engage in snowdrift games with their active neighbors and dynamically adjust their participation states according to the difference between their realized payoffs and aspiration levels. At the decision-making level, enterprises employ Q-learning to adaptively select between conformist and payoff-driven modes according to the local active environment. Active enterprises adopting different decision-making modes then update their resource-sharing strategies according to different rules. Conformist enterprises tend to follow the resource-sharing behavior of their neighbors, whereas payoff-driven enterprises update their strategies based on payoff differences between themselves and neighboring enterprises. Through these coupled mechanisms, participation states, decision-making modes, and resource-sharing strategies interact through feedback and co-evolve on the enterprise resource-sharing network.

The remainder of this paper is organized as follows. Section~\ref{Model} introduces the proposed resource-sharing game model with dynamic participation and adaptive decision-making. Section~\ref{Simulation} presents the simulation results and analyzes the underlying evolutionary mechanisms. Finally, Section~\ref{Conclusion} summarizes the main findings and discusses possible directions for future research.

\section{Model} \label{Model}

In this section, we introduce the proposed model in detail. We first describe the enterprise resource-sharing network and the game interactions among enterprises. We then introduce an aspiration-based mechanism to characterize the dynamic participation of enterprises in resource-sharing activities. Subsequently, two decision-making modes, namely payoff-driven and conformist modes, are introduced to determine how enterprises update their resource-sharing strategies~\cite{szolnoki_njp18b, zhang_lm_pa21, wang_p_csf24}. Next, we present a Q-learning framework through which enterprises adaptively switch between these two decision-making modes.

\subsection{Resource-sharing network and game model}

We consider a resource-sharing network consisting of $N$ enterprises, represented by an undirected complex network $\mathcal{G}=(\mathcal{V},\mathcal{E})$, where each node $i\in\mathcal{V}$ represents an enterprise and each edge $(i,j)\in\mathcal{E}$ represents a potential resource-sharing relationship between enterprises $i$ and $j$. The existence of a network connection, however, does not necessarily imply that connected enterprises participate in resource sharing at all times. To characterize dynamic participation, each enterprise $i$ is assigned a participation state $a_i\in\{0,1\}$, where $a_i=1$ indicates that the enterprise $i$ is active and participates in the current resource-sharing game, 
while
$a_i=0$ indicates that it is dormant and temporarily does not participate.

Each active enterprise interacts with all 
its active neighbors and adopts one of two resource-sharing strategies, i.e., cooperation $C$ or defection $D$. Strategy $C$ represents resource sharing, whereby an enterprise provides available resources to its interaction partners, whereas strategy $D$ represents non-sharing, whereby an enterprise retains its own resources while potentially benefiting from resources shared by others. Resource sharing generally involves both costs and benefits. When two enterprises choose $C$, they mutually benefit from resource sharing. When a $C$ enterprise interacts with a $D$ enterprise, the latter can benefit from the resources provided by the former without incurring a corresponding sharing cost, whereas the former receives a lower payoff because it bears the cost of sharing. When both enterprises choose $D$, neither 
receives
additional benefits from resource sharing. This payoff structure corresponds to the snowdrift game. Accordingly, the payoff matrix is given by
\begin{equation}\label{Payoff matrix}
M=
\left(
\begin{array}{cc}
1 & 1-r \\
1+r & 0  \\
\end{array}
\right),
\end{equation}
where $r\in[0,1]$ characterizes the relative cost of resource sharing with respect to its benefit.

At each time step, an active enterprise accumulates payoffs by playing the resource-sharing game with all its active neighbors. Although a dormant enterprise does not participate in the resource-sharing game, it can still obtain a baseline payoff from its own resources and existing network relationships. Therefore, the payoff of enterprise $i$ is defined as
\begin{equation}\label{Payoff calculation}
\Pi_i =\begin{cases}
	\sum \limits_{j\in N_i}{a_j s_{i}^{T} M s_j}, &\text{if } a_i=1,\\
	k_i\beta, &\text{if } a_i=0,\\
\end{cases}
\end{equation}
where $N_i$ denotes the set of neighbors of the enterprise $i$ and $k_i=|N_i|$ is its degree. The strategy of the enterprise $i$ is represented by the vector $s_i$, with $s_i=[1,0]^T$ corresponding to strategy $C$ and $s_i=[0,1]^T$ corresponding to strategy $D$. Here, $\beta$ represents the baseline payoff per network connection for a dormant enterprise. We note that the underlying network remains static and that temporarily suspending direct resource-sharing interactions does not eliminate an enterprise's existing network relationships. Accordingly, the dormant payoff depends on the total degree, rather than solely on the number of currently active neighbors.

\subsection{Dynamic participation mechanism with an aspiration-based learning rule}

The participation state of an enterprise in resource-sharing activities is not fixed, but evolves dynamically according to the discrepancy between its actual payoff and aspiration level. Specifically, each enterprise is assigned an aspiration level $A$, which represents the payoff it expects to obtain. When the actual payoff of the enterprise $i$ falls below its aspiration level, it becomes dissatisfied with its current participation state and is therefore more likely to switch between the active and dormant states. The probability that the enterprise $i$ changes its participation state is given by
\begin{equation}
P_i=\frac{1}{1+\exp\left[-\lambda\left(A-\Pi_i\right)\right]},
\end{equation}
where $\lambda>0$ denotes the sensitivity of an enterprise to the difference between its aspiration level and the actual payoff. A larger $\lambda$ results in a sharper response to the payoff-aspiration difference: when $\Pi_i<A$, the probability of switching increases more rapidly as the payoff falls below the aspiration level, whereas when $\Pi_i>A$, the probability of switching decreases accordingly.

The participation state of the enterprise $i$ is then updated according to
\begin{equation}
a_i(t+1)=
\begin{cases}
1-a_i(t), & \text{with probability } P_i, \\
a_i(t), & \text{with probability } 1-P_i.  \\
\end{cases}
\end{equation}

We emphasize that this mechanism describes bidirectional state transitions rather than a one-way withdrawal process. Thus, both active and dormant enterprises can switch their participation states depending on the relationship between their actual payoffs and aspiration levels. This mechanism enables enterprises to dynamically enter or withdraw from resource-sharing activities and generates 
endogenous feedback between payoff performance and participation dynamics.

\subsection{Strategic evolution under two decision-making modes}

After enterprises obtain their payoffs from the resource-sharing game, active enterprises update their resource-sharing strategies, whereas dormant enterprises retain their current strategies. Unlike conventional evolutionary game models in which all individuals follow the same strategy-update rule, we consider two distinct decision-making modes for active enterprises, including payoff-driven and conformist modes, both of which are widely observed in reality. The former represents payoff-based strategy imitation, while the latter describes conformity to the locally prevailing strategy. Specifically, for an active payoff-driven enterprise $i$, the enterprise first randomly selects an enterprise $j$ from its active neighbors as a reference and decides whether to imitate its resource-sharing strategy based on the difference in their payoffs. The probability that 
a
payoff-driven enterprise $i$ adopts the strategy of 
an
active neighbor $j$ is given as follows
\begin{equation}
W_{\mathrm{Pay}}(s_{i}\leftarrow s_{j})=\frac{1}{1+\exp[(\Pi_i-\Pi_j)/\kappa]},
\end{equation}
where $\kappa>0$ denotes the noise intensity in the strategy-update process and characterizes 
the
stochasticity in enterprise decision-making. When $\Pi_j>\Pi_i$, the enterprise $i$ adopts the strategy of 
the
enterprise $j$ with a probability greater than $0.5$, indicating a tendency to imitate neighbors with higher payoffs. 
In contrast,
when $\Pi_j<\Pi_i$, 
the
enterprise $i$ is more likely to retain its current strategy.

For active conformist enterprises, strategy updates do not depend directly on payoffs but are primarily determined by the strategy distribution among their active neighbors~\cite{szolnoki_rsif15}. 
The
conformist enterprise $i$ is more likely to choose the strategy used by the majority of its active neighbors, and the probability that it changes its current strategy can be expressed as
\begin{equation}
W_{\mathrm{Con}}(s_{i}\,\,\mathrm{change})=\frac{1}{1+\exp[(N_i^{\mathrm{same}}-\frac{k_i^A}{2}) / \kappa]},
\end{equation}
where $k_i^A$ represents the number of active enterprises in the neighborhood of enterprise $i$, and $N_i^{\mathrm{same}}$ indicates the number of active neighbors that share the same strategy as enterprise $i$, defined as
\begin{equation}
N_i^{\mathrm{same}}=\sum_{j\in N_i}a_j\delta(s_i,s_j),
\end{equation}
where $\delta(s_i,s_j)=1$ if $s_i=s_j$ and $\delta(s_i,s_j)=0$ otherwise. When the current strategy of the enterprise $i$ is supported by the majority of its active neighbors, i.e., $N_i^{\mathrm{same}}>k_i^A/2$, the probability of changing its strategy is less than $0.5$, and 
therefore the enterprise
more likely to retain its current strategy. By contrast, when its current strategy is in the local minority, the switching probability exceeds $0.5$, causing the enterprise to preferentially align its strategy with the local majority.

\subsection{Adaptive decision-making mode selection based on Q-learning}

In reality, the decision-making mode of an enterprise is not fixed during the evolutionary process but can be adaptively adjusted according to its accumulated experience. To characterize this adaptive behavior, we introduce Q-learning, which enables enterprises to learn which decision-making mode is more suitable
in
different local resource-sharing environments.

Each enterprise $i$ maintains a Q-table $Q_i(S,A)$ that estimates the expected long-term return associated with selecting action $A$ in state $S$. The state of enterprise $i$ is defined by the number of its active neighbors. Thus, for an enterprise with degree $k_i$, the state space is $\mathcal{S}_i=\{0,1,\ldots,k_i\}$. The action space is defined as $\mathcal{A}=\{\mathrm{Pay},\mathrm{Con}\}$, where $\mathrm{Pay}$ and $\mathrm{Con}$ denote the payoff-driven and conformist decision-making modes, respectively. The Q-table of each enterprise is updated according to the following rule
\begin{equation}  \label{Q-learning}
\begin{aligned}
Q_i^{t+1}(S,A)=&Q_i^t(S,A)+\alpha [r_i^t(S,A)\\
&+\gamma\max_{A'\in\mathcal{A}}Q_i^t(S',A')-Q_i^t(S,A)],
\end{aligned}
\end{equation}
where $\alpha$ indicates the learning rate leveraged to control the update rate of the Q-value, $\gamma$ is the discount factor employed to control the value of future reward, and $S' \in \mathcal{S}$ represents the new state after taking action $A$ 
in
state $S$. To take into account both an enterprise's own payoff and the local resource-sharing environment, the immediate reward $r_i^t$ of enterprise $i$ at time step $t$ is defined as
\begin{equation} \label{reward}
r_i^t=\omega\times\Pi_i+(1-\omega)\times\frac{1}{k_i}\sum_{j\in N_i}\Pi_j,
\end{equation}
where $\omega\in[0,1]$ controls the relative importance assigned to individual and neighborhood payoffs. A 
higher
value of $\omega$ indicates that the enterprise places greater emphasis on its own payoff, whereas a 
lower
value means greater consideration of the payoff performance of neighboring enterprises. This formulation allows different degrees of individual and local social concerns to be incorporated into the learning process.

After updating its Q-table, the enterprise $i$ selects its decision-making mode for the next time step according to an $\epsilon$-greedy policy. With probability $\epsilon$, the enterprise explores by randomly selecting one of the two decision-making modes, whereas with probability $1-\epsilon$, it exploits its accumulated experience by choosing the action with the highest Q-value. Therefore, the choice of decision-making mode for enterprise $i$ can be expressed as
\begin{equation}
\tau_i=
\begin{cases}
\mathrm{Pay}, & \text{with probability } \epsilon/2, \\
\mathrm{Con}, & \text{with probability } \epsilon/2, \\
\displaystyle\arg\max_{A\in\mathcal{A}}Q_i(S,A), & \text{with probability } 1-\epsilon. \\
\end{cases}
\end{equation}
Through repeated interaction and learning, enterprises can explore alternative decision-making modes and gradually adapt their decision-making modes to the local resource-sharing environment.

Overall, the model proposed in this paper couples three evolutionary processes, including dynamic participation, resource-sharing strategy evolution, and adaptive decision-making. Enterprise participation is governed by the aspiration-based switching mechanism, resource-sharing strategies evolve according to either a payoff-driven or a conformist updating rule, and the decision-making mode is adaptively selected through Q-learning. These coupled processes jointly determine the evolution of cooperation in the enterprise resource-sharing network.

\section{Simulation results and analysis} \label{Simulation}

In this section, we perform extensive numerical simulations to investigate the coupled evolution of enterprise participation, decision-making modes, and resource-sharing strategies in resource-sharing networks. Unless otherwise specified, the underlying network is modeled as a Watts--Strogatz (WS) small-world network with $N=5000$ nodes, mean degree $k=4$, and rewiring probability $p=0.15$, motivated by the prevalence of small-world structures in many real-world systems. The default Q-learning parameters are $\epsilon=0.1$, $\gamma=0.5$, and $\alpha=0.9$. The default aspiration level and participation-switching sensitivity are set to $A=1.5$ and $\lambda=2$, respectively, and the payoff weight is $\omega = 0.5$. The default payoff parameters are $r=0.3$ and $\beta=0.4$. These values serve as baseline settings and are varied in the corresponding simulations when their effects on the system dynamics are investigated. Initially, each enterprise independently adopts the active or dormant state with equal probability. Similarly, the resource-sharing strategy ($C$ or $D$) and the decision-making mode (payoff-driven or conformist) are each initialized with probability $0.5$.

Our simulations show that the fraction of active enterprises, denoted by $f_a$, and the fraction of cooperative enterprises, denoted by $f_c$, generally reach stationary levels after approximately $t=1000$ time steps. The cooperation fraction $f_c$ is defined as the fraction of cooperative enterprises in the entire population. This population-level measure captures the overall prevalence of cooperative strategies, regardless of enterprises' current participation states. Although dormant enterprises do not participate in resource-sharing interactions while dormant, they retain their strategies. When a dormant cooperator becomes active again, it can immediately contribute to active cooperative interactions and influence subsequent strategy evolution. Accordingly, $f_c$ measures the fraction of cooperative enterprises in both active and dormant states. For the fraction of payoff-driven enterprises, it fluctuates around $0.5$ without exhibiting persistent dominance over the conformist mode. Therefore, the following analysis focuses primarily on the participation of enterprises and cooperation in resource-sharing. This does not imply that decision-making modes have no influence on evolutionary dynamics; rather, under the parameter ranges considered here, the two modes remain dynamically balanced at the population level. To reduce the influence of temporal fluctuations, each simulation is run for $5000$ time steps, and the reported quantities are averaged over the final $500$ steps after the system has reached a stationary regime. In addition, each parameter setting is averaged over $10$ independent realizations.

Based on these settings, we systematically examine the effects of the main model parameters and mechanisms on enterprise participation and cooperation. In particular, we first investigate how the payoff parameters $\beta$ and $r$ affect the stationary active fraction $f_a$ and the cooperation fraction $f_c$. We then examine the effects of the aspiration level $A$ and the payoff weight $\omega$ on these two quantities. To evaluate the robustness of the results with respect to reinforcement-learning parameters, we further analyze the effects of the exploration rate $\epsilon$ and the learning rate $\alpha$ on $f_a$ and $f_c$ using $\epsilon$--$\alpha$ parameters. We subsequently investigate the combined effects of the participation-switching sensitivity $\lambda$ and the payoff parameter $r$ on the evolutionary dynamics of four categories of enterprises: cooperative conformist enterprises (CC), cooperative payoff-driven enterprises (CP), defective conformist enterprises (DC), and defective payoff-driven enterprises (DP). In addition, spatial snapshots on a $100\times100$ square lattice (SL) with periodic boundary conditions and von~Neumann neighborhood are used to illustrate the evolution of participation states and resource-sharing strategies under different aspiration levels. Furthermore, we compare the dependence of $f_c$ on $r$ in WS and SL networks, both with and without the dynamic participation mechanism, to assess how participation dynamics and network structure jointly influence the evolution of cooperation. To distinguish the contribution of adaptive mode selection from the effect of maintaining a mixture of decision-making modes, we conduct an additional ablation study comparing the original Q-learning model with a fixed 50:50 mixture and two homogeneous baselines. Finally, we perform simulations on WS and SL networks of different sizes to determine whether the observed cooperation patterns persist as the system size changes.

\subsection{The impact of payoff parameters on enterprise participation and cooperative behavior}

\begin{center}
\begin{figure}[htbp]
\centering
\subfigure[$f_a$]{
\includegraphics[scale=0.28]{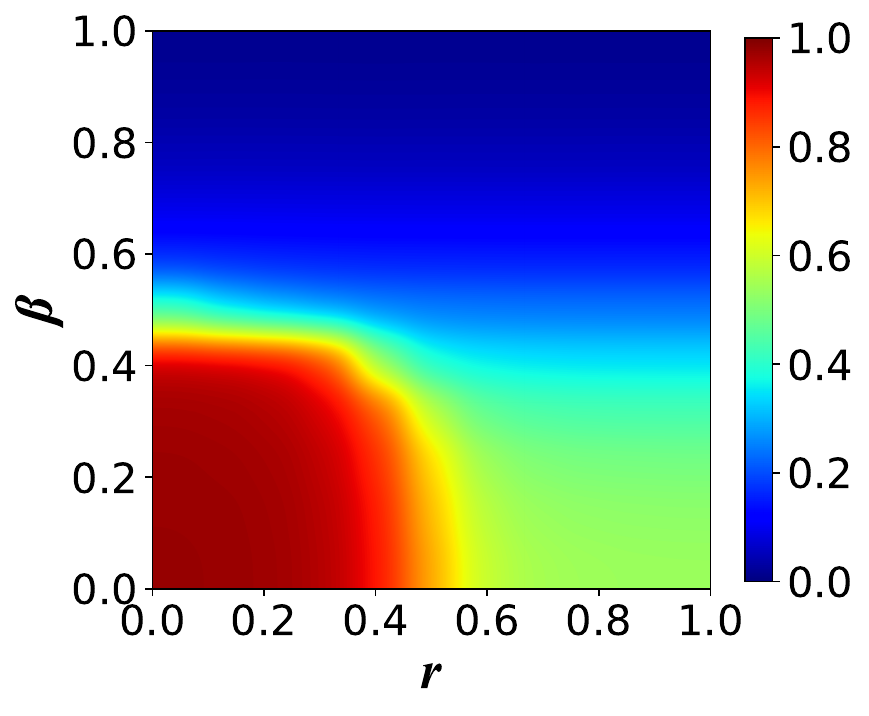}
\label{fa_beta_r_WS}}
\subfigure[$f_c$]{
\includegraphics[scale=0.28]{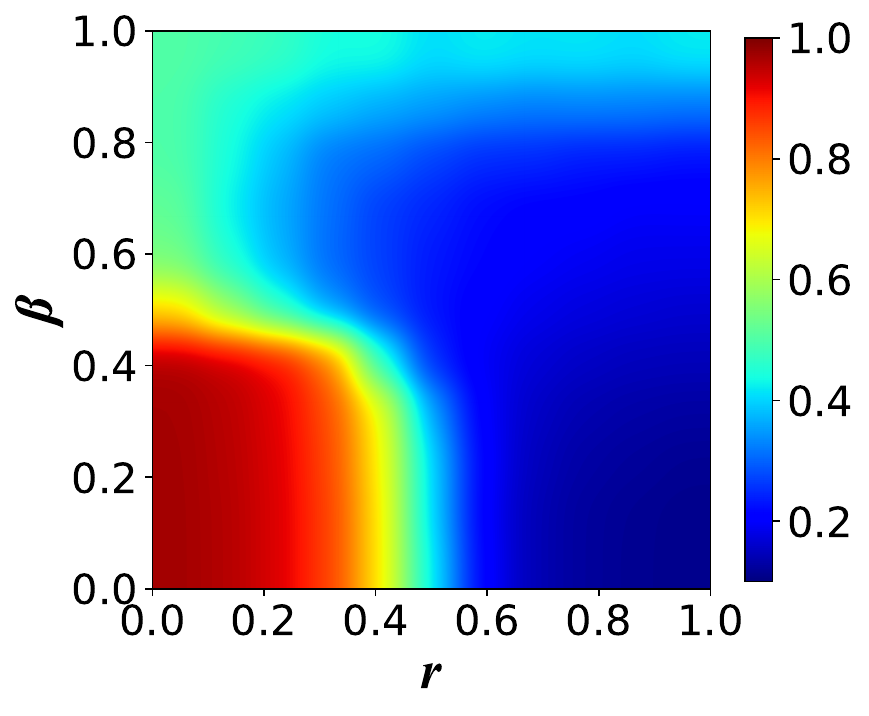}
\label{fc_beta_r_WS}}
\caption{\textbf{Effects of the payoff parameters $\beta$ and $r$ on enterprise participation and cooperation.} Panels~(a) and (b) show the stationary active fraction $f_a$ and cooperation fraction $f_c$, respectively. In both panels, $r$ and $\beta$ vary over the interval $[0,1]$. The active fraction generally decreases as either parameter increases, whereas the effect of $\beta$ on cooperation depends on the value of $r$. In particular, increasing $\beta$ suppresses cooperation at small $r$ but can promote it at large $r$, highlighting the interplay between participation dynamics and cooperation evolution.}
\label{beta_r}
\end{figure}
\end{center}

In this subsection, we investigate the joint effects of the payoff parameters $\beta$ and $r$ on enterprise participation and resource-sharing cooperation. To characterize their combined influence on the stationary evolutionary outcomes, we systematically vary $\beta$ and $r$ over the interval $[0,1]$ and measure the stationary active fraction $f_a$ and the cooperation fraction $f_c$. The results are shown in Fig.~\ref{beta_r}.

As shown in Fig.~\ref{fa_beta_r_WS}, the fraction of active enterprises generally decreases as either $\beta$ or $r$ increases. When both parameters are small, the system maintains a relatively high level of participation. As $\beta$ increases, however, $f_a$ declines markedly, indicating that a higher baseline payoff in the dormant state makes continued participation in resource-sharing activities less attractive. Increasing $r$ similarly suppresses participation by making resource sharing less favorable in the underlying snowdrift game.

The dependence of cooperation on $\beta$ and $r$, shown in Fig.~\ref{fc_beta_r_WS}, is more complex. In particular, the effect of $\beta$ on $f_c$ depends strongly on the value of $r$. For small $r$, increasing $\beta$ leads to a decrease in $f_c$, whereas for sufficiently large $r$, the opposite trend emerges and $f_c$ increases with $\beta$. This non-monotonic response indicates a clear interaction between the two payoff parameters rather than two independent effects.

A possible explanation lies in the coupling between participation dynamics and strategy evolution. The parameter $\beta$ determines the baseline payoff received by dormant enterprises and therefore directly affects their tendency to remain dormant or return to the resource-sharing game. By contrast, $r$ regulates the payoff differences among different combinations of strategies in the snowdrift game and thus influences the competitive balance between cooperation and defection. When $r$ is small, increasing $\beta$ mainly enhances the relative attractiveness of dormancy, thereby suppressing both participation and cooperation. At larger $r$, however, changes in participation reshape the effective interaction environment experienced by active enterprises. This altered participation structure can modify the relative evolutionary advantages of cooperative and defective strategies, leading to an increase in the observed cooperation fraction even as the overall active fraction decreases.

\subsection{The impact of aspiration and payoff weight on enterprise participation and cooperative behavior}

\begin{center}
\begin{figure}[htbp]
\centering
\subfigure[$f_a$]{
\includegraphics[scale=0.28]{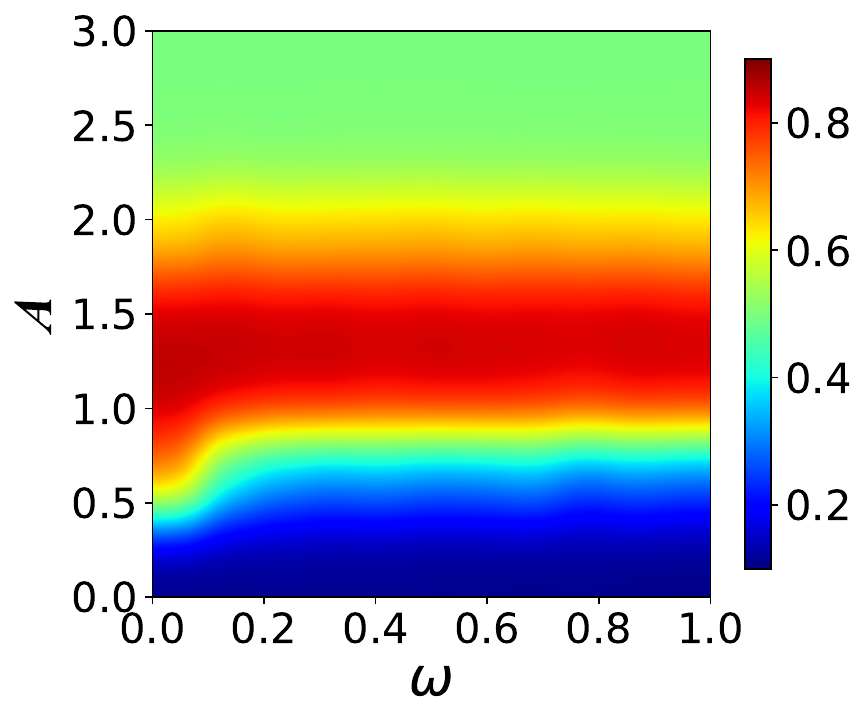}
\label{fa_aspir_omega_WS}}
\subfigure[$f_c$]{
\includegraphics[scale=0.28]{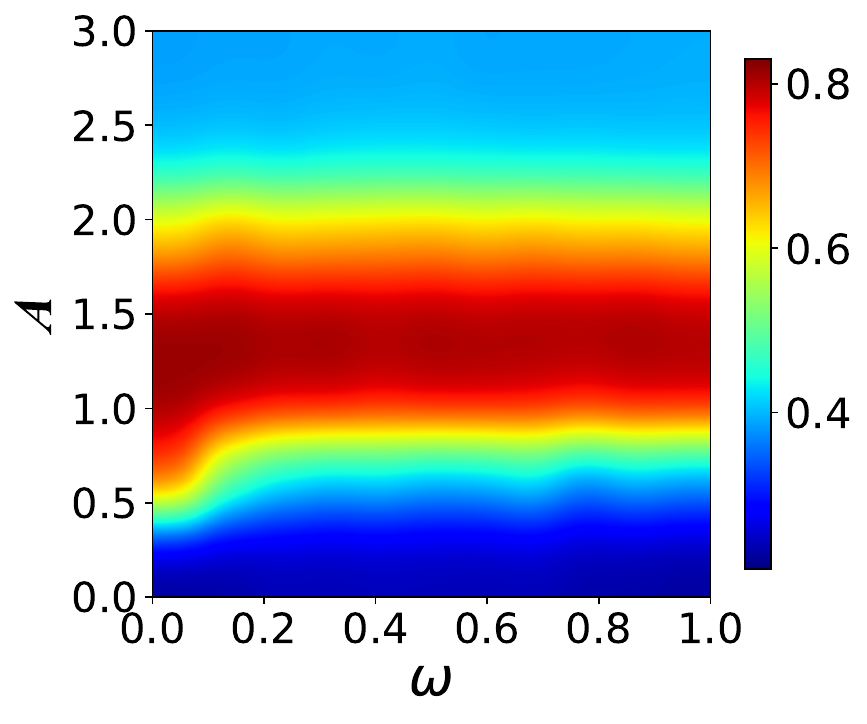}
\label{fc_aspir_omega_WS}}
\caption{\textbf{Heatmaps of the final proportion of active enterprises and cooperative enterprises in dependence of aspiration level $A$ and payoff weight $\omega$.} Panel~(a) shows the final proportion of active enterprises $f_a$. Panel~(b) corresponds to the final frequency of cooperative enterprises $f_c$. Both $f_a$ and $f_c$ exhibit a non-monotonic dependence on $A$, reaching relatively high values at intermediate aspiration levels. In comparison, $\omega$ has a weaker influence on both quantities, and the overall non-monotonic trend with respect to $A$ persists across the examined values of $\omega$.}
\label{aspir_omega}
\end{figure}
\end{center}

Next, we investigate the effects of the aspiration level $A$ and payoff weight $\omega$ on enterprise participation and cooperation. Figs. \ref{fa_aspir_omega_WS} and \ref{fc_aspir_omega_WS} show the stationary active fraction $f_a$ and cooperation fraction $f_c$, respectively, for different combinations of $A$ and $\omega$.

From Fig.~\ref{aspir_omega}, we can see that the aspiration level exhibits a pronounced non-monotonic effect on both $f_a$ and $f_c$. As $A$ increases from a low value, both quantities initially increase, reach relatively high levels at intermediate aspiration values, and then decrease as $A$ increases further. This indicates that neither excessively low nor high aspiration levels are favorable for sustaining enterprise participation and cooperation, whereas an intermediate range of aspiration promotes both.

This non-monotonic behavior can be understood from the aspiration-based participation mechanism. When $A$ is relatively low, enterprises are easily satisfied with their current payoffs, resulting in a low tendency to switch participation states. Consequently, the dynamic participation mechanism plays only a limited role in reorganizing the effective interaction structure of the network. As $A$ increases to an intermediate level, enterprises become more responsive to unfavorable payoff conditions, and appropriate switching between active and dormant states enables them to adapt their participation to changes in the local environment. This dynamic adjustment creates more favorable conditions for maintaining participation and cooperation. When $A$ becomes excessively high, however, enterprises are more likely to remain dissatisfied with their realized payoffs, leading to frequent switching between active and dormant states. Such excessive state switching disrupts persistent interactions among enterprises and ultimately reduces both $f_a$ and $f_c$.

Compared to aspiration, the payoff weight $\omega$ has a relatively weaker influence on the stationary values of $f_a$ and $f_c$, although some interaction between $A$ and $\omega$ can still be observed. According to Eq.~(\ref{reward}), $\omega$ determines the relative importance assigned to an enterprise's own payoff and the average payoff of its neighbors in the reinforcement-learning reward. Changes in $\omega$ therefore affect the learned preference for payoff-driven and conformist decision-making modes, which in turn indirectly influence strategy evolution and participation dynamics. As a result, the aspiration range associated with relatively high participation and cooperation varies moderately with $\omega$, while the overall non-monotonic dependence on $A$ remains robust.

\subsection{The impact of Q-learning parameters on enterprise participation and cooperative behavior}

\begin{center}
\begin{figure}[htbp]
\centering
\subfigure[$f_a$]{
\includegraphics[scale=0.27]{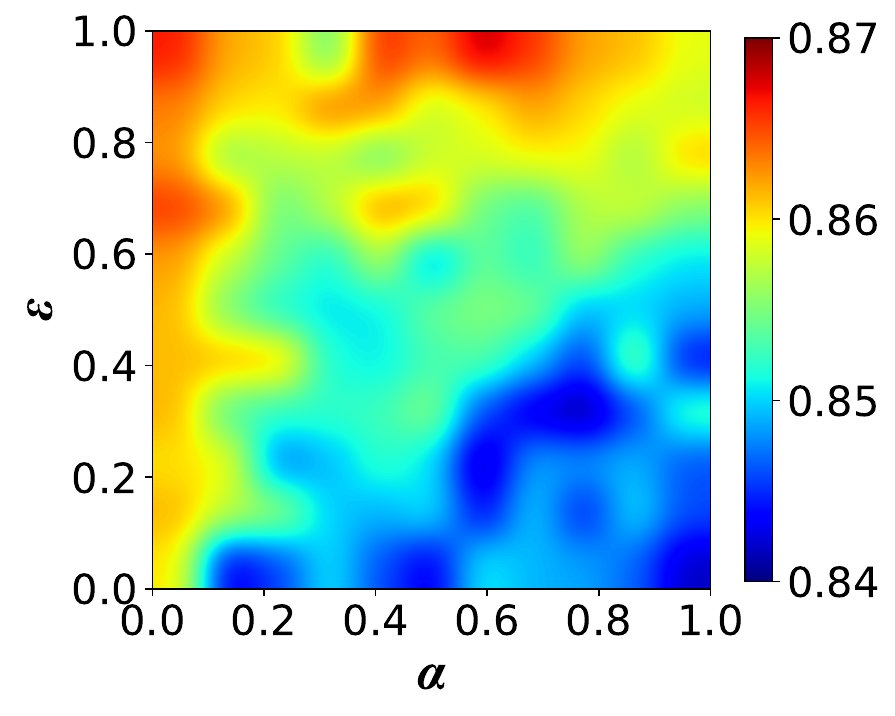}
\label{fa_epsilon_alpha_WS}}
\subfigure[$f_c$]{
\includegraphics[scale=0.27]{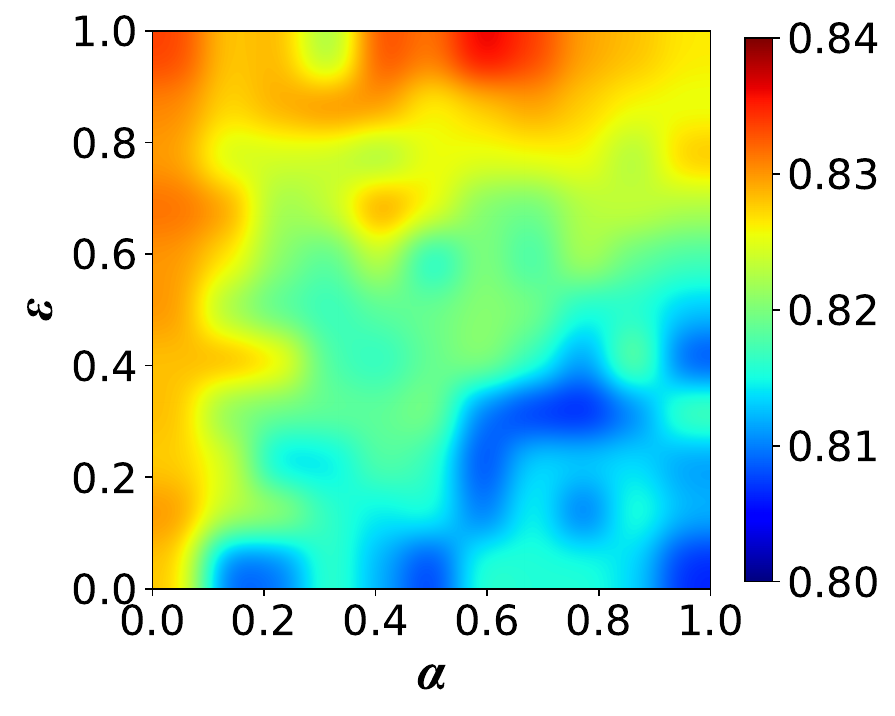}
\label{fc_epsilon_alpha_WS}}
\caption{\textbf{Effects of the Q-learning parameters $\epsilon$ and $\alpha$ on enterprise participation and cooperation.} Panels~(a) and (b) show heatmaps of the stationary active fraction $f_a$ and cooperation fraction $f_c$, respectively, as functions of the exploration rate $\epsilon$ and learning rate $\alpha$. Both quantities vary only slightly across the examined parameter range, indicating that the macroscopic levels of participation and cooperation are relatively robust to changes in the Q-learning parameters.}
\label{epsilon_alpha}
\end{figure}
\end{center}

To assess the sensitivity of the evolutionary outcomes to the Q-learning parameters, we investigate the effects of the exploration rate $\epsilon$ and learning rate $\alpha$ on the stationary active fraction $f_a$ and cooperation fraction $f_c$. As shown in Fig.~\ref{epsilon_alpha}, both quantities remain relatively stable over the parameter range $\epsilon,\alpha\in[0,1]$. Specifically, $f_a$ varies approximately between $0.84$ and $0.87$, whereas $f_c$ remains within the range of approximately $0.80$--$0.84$.

A weak gradient can nevertheless be observed across the heatmaps. The values of both $f_a$ and $f_c$ are slightly higher in the region characterized by larger $\epsilon$ and smaller $\alpha$ than in the region with smaller $\epsilon$ and larger $\alpha$. However, the overall variation remains limited. Since $\epsilon$ determines the extent to which enterprises explore alternative decision-making modes, whereas $\alpha$ controls the weight assigned to newly acquired information during Q-value updating, these results indicate that changes in the learning dynamics do not substantially alter the macroscopic levels of participation and cooperation. The qualitative evolutionary outcomes of the model are therefore relatively insensitive to the specific choices of $\epsilon$ and $\alpha$ within the parameter range considered.

\subsection{Evolution of four enterprise types}

\begin{center}
\begin{figure}[htbp]
\centering
\includegraphics[scale=0.3]{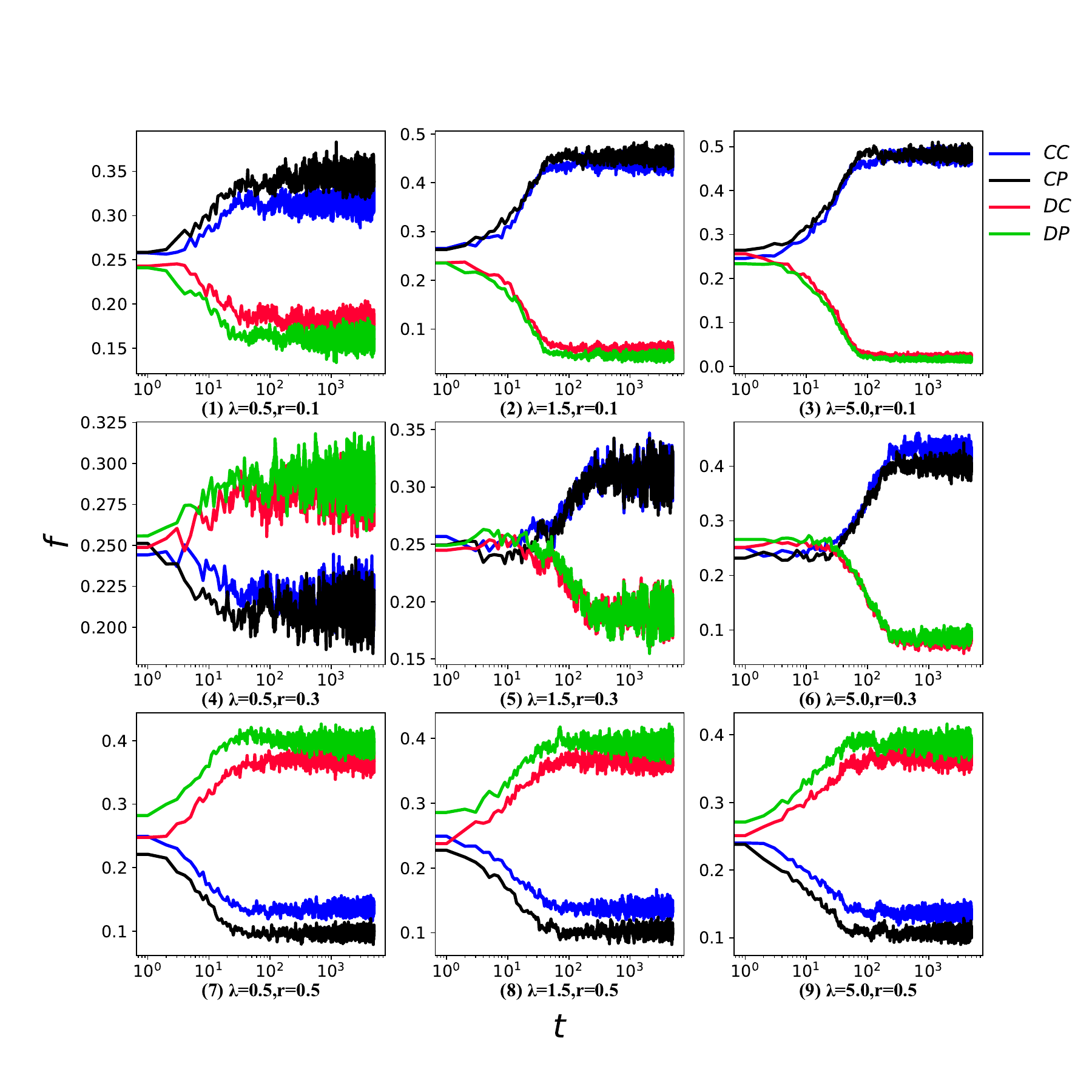}
\caption{\textbf{Temporal evolution of the four enterprise categories under different combinations of $\lambda$ and $r$.} The $x$-axis denotes evolutionary time on a logarithmic scale, and the $y$-axis represents the population fraction of each enterprise category. The four categories are cooperative conformist enterprises (CC), cooperative payoff-driven enterprises (CP), defective conformist enterprises (DC), and defective payoff-driven enterprises (DP). Different panels correspond to different parameter combinations, with $\lambda\in\{0.5,1.5,5.0\}$ and $r\in\{0.1,0.3,0.5\}$. As $r$ increases, the combined fraction of cooperative enterprises generally decreases, while the influence of $\lambda$ on the stationary population composition depends on the value of $r$ and becomes weaker at larger $r$.}
\label{evolutionary curves}
\end{figure}
\end{center}

In this subsection, we further explore how the participation-switching sensitivity $\lambda$ and payoff parameter $r$ affect the coupled evolution of decision-making modes and resource-sharing strategies. Fig.~\ref{evolutionary curves} shows the temporal evolution of four enterprise categories: cooperative conformist enterprises (CC), cooperative payoff-driven enterprises (CP), defective conformist enterprises (DC), and defective payoff-driven enterprises (DP), under different combinations of $\lambda$ and $r$. Since the population fractions of these four categories approach stationary levels relatively rapidly, the evolutionary time on the $x$-axis is displayed on a logarithmic scale to better resolve the transient dynamics at early stages.

As displayed in Fig.~\ref{evolutionary curves}, the fractions of the four enterprise categories undergo pronounced transient adjustments before approaching relatively stable levels. Overall, increasing $r$ substantially changes the population composition. When $r$ is small, the cooperative categories CC and CP account for relatively large fractions of the population. As $r$ increases, their combined proportion decreases, accompanied by a corresponding increase in the prevalence of defective enterprises. This behavior reflects the role of $r$ in reshaping the payoff differences between cooperative and defective strategies and, consequently, the evolutionary balance among the four enterprise categories.

The parameter $\lambda$, by contrast, mainly regulates how sensitively enterprises adjust their participation states in response to the discrepancy between realized payoffs and aspiration levels. Increasing $\lambda$ therefore modifies the transient evolutionary trajectories of CC, CP, DC, and DP by strengthening the responsiveness of participation-state switching. For relatively small values of $r$, such as $r=0.1$ in the first row and $r=0.3$ in the second row, increasing $\lambda$ is associated with a higher prevalence of cooperative enterprises. This suggests that stronger participation adjustment can reshape the effective payoff environment and thereby indirectly influence both strategy evolution and the learned decision-making modes. When $r=0.5$ in the third row, however, the stationary population composition is much less sensitive to changes in $\lambda$, indicating that the influence of the payoff structure becomes dominant under less favorable conditions for cooperation.

We note that the four enterprise categories are not independent evolutionary groups. Each category combines two behavioral dimensions: the resource-sharing strategy ($C$ or $D$) and the decision-making mode (conformist or payoff-driven). Changes in either dimension can therefore feed back into the other. A strategy change modifies enterprise payoffs, which can subsequently affect participation-state switching and Q-learning-based mode selection. Conversely, changes in the selected decision-making mode alter the strategy-update rule and thereby influence the subsequent evolution of cooperation. The observed population composition thus emerges from the coupled feedback among participation dynamics, strategy evolution, and adaptive decision-making.

\subsection{Spatial evolution of enterprise participation and behavioral patterns under different aspiration levels}

\begin{center}
\begin{figure}[htbp]
\centering
\includegraphics[scale=0.28]{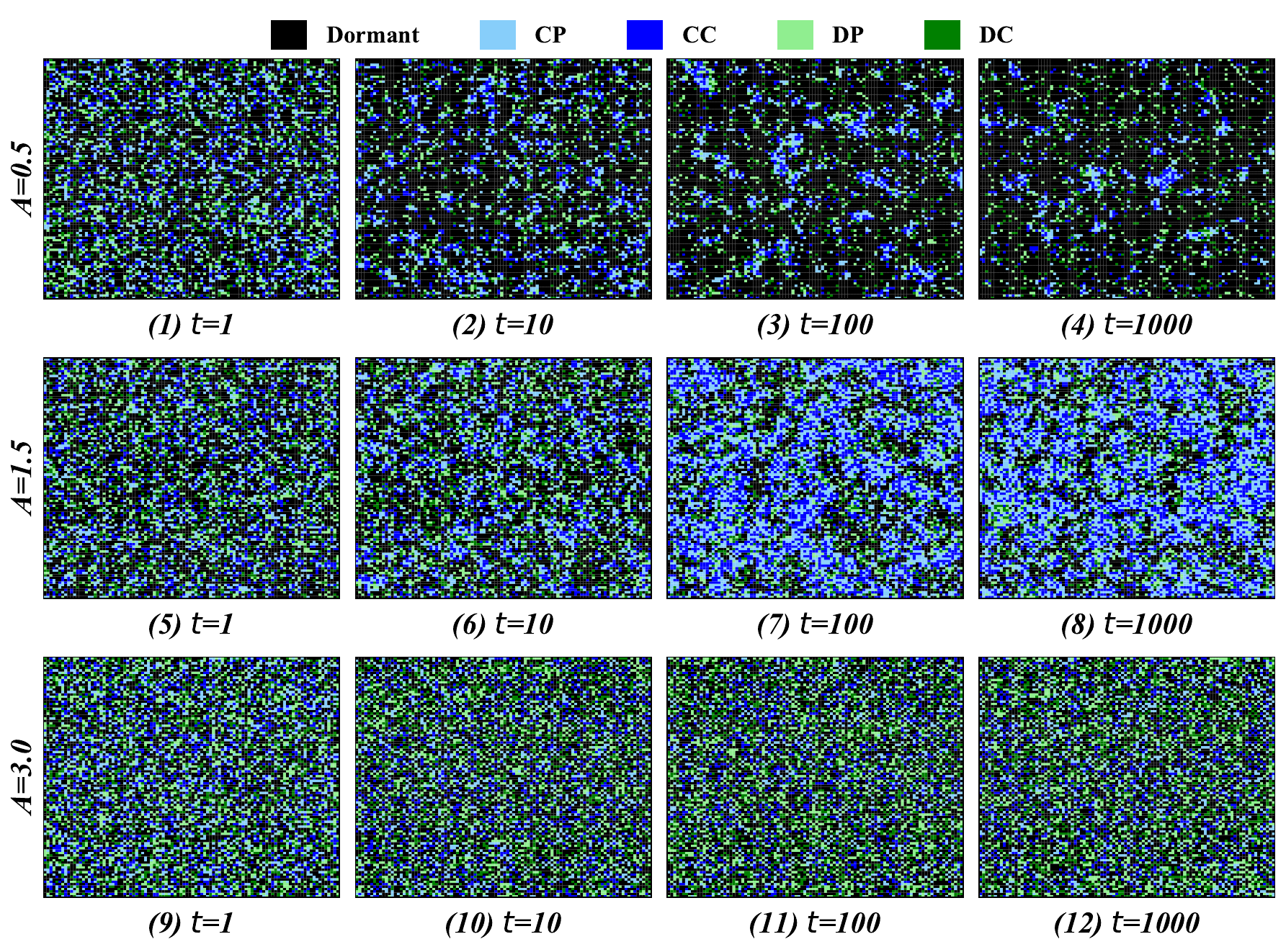}
\caption{\textbf{Evolution of spatial distribution of enterprises on a $100\times100$ square lattice under different aspiration levels $A$.} Different colors represent different states of enterprises.
Black denotes dormant enterprises, while light blue, dark blue, light green, and dark green represent cooperative payoff-driven (CP), cooperative conformist (CC), defective payoff-driven (DP), and defective conformist (DC) enterprises, respectively. Rows correspond to $A=0.5$, $1.5$, and $3.0$, while columns correspond to $t=1$, $10$, $100$, and $1000$. An intermediate aspiration level sustains a higher level of participation and more extensive cooperative clusters, whereas both lower and higher aspiration levels result in less prevalent active and cooperative regions.}
\label{snapshots}
\end{figure}
\end{center}

Subsequently, we illustrate the spatial evolution of enterprise participation, decision-making modes, and resource-sharing strategies. We perform simulations on a $100\times100$ square lattice with periodic boundary conditions and von~Neumann neighborhood under different aspiration levels $A$. Spatial configurations are recorded at $t=1$, $10$, $100$, and $1000$, as shown in Fig.~\ref{snapshots}. Dormant enterprises are represented in black, whereas active enterprises are classified as cooperative payoff-driven (CP), cooperative conformists (CC), defective payoff-driven (DP), and defective conformists (DC) enterprises, which are demonstrated in light blue, dark blue, light green, and dark green, respectively. Thus, CC and CP represent enterprises that are both cooperative and currently participating in resource-sharing interactions, whereas DC and DP represent active defectors.

From Fig.~\ref{snapshots}, it is evident that enterprises of different categories are initially distributed relatively randomly across the lattice. As the system evolves, their participation states, decision-making modes, and resource-sharing strategies continuously adjust, gradually giving rise to spatially clustered structures. The stationary level of participation exhibits a clear non-monotonic dependence on the aspiration level $A$. When $A=0.5$ in the first row, the active fraction is relatively low and eventually approaches approximately $0.2$, leaving most of the lattice occupied by dormant enterprises. At the intermediate aspiration level $A=1.5$ in the second row, the active fraction increases markedly to approximately $0.7$, resulting in substantially larger active regions. When the aspiration level is further increased to $A=3.0$ in the third row, the active fraction decreases again to approximately $0.5$. These spatial patterns indicate that an intermediate aspiration level is more favorable for sustaining enterprise participation, whereas both relatively low and high aspiration levels reduce the prevalence of active enterprises.

A similar nonmonotonic pattern can be observed in the spatial distribution of active cooperators. At $A=1.5$, active cooperative enterprises are more prevalent and form larger, more extensive spatial clusters. By contrast, at $A=0.5$ and $A=3.0$, active cooperators are less abundant, and large clusters of active cooperation are less likely to emerge and persist. These observations suggest that an intermediate aspiration level not only supports enterprise participation but also creates more favorable conditions for the formation and persistence of spatial clusters of active cooperators.

The spatial results are consistent with the heatmap analysis shown in Fig.~\ref{aspir_omega}, where both $f_a$ and $f_c$ exhibit non-monotonic dependence on aspiration. The lattice simulations further demonstrate that aspiration affects not only the macroscopic levels of participation and cooperation but also the spatial organization of enterprise states. Thus, aspiration plays an important nonlinear role in shaping both the global and local evolutionary patterns of enterprise resource sharing.

\subsection{The influence of dynamic participation and Q-learning on the evolution of cooperation}

\begin{center}
\begin{figure}[htbp]
\centering
\includegraphics[scale=0.26]{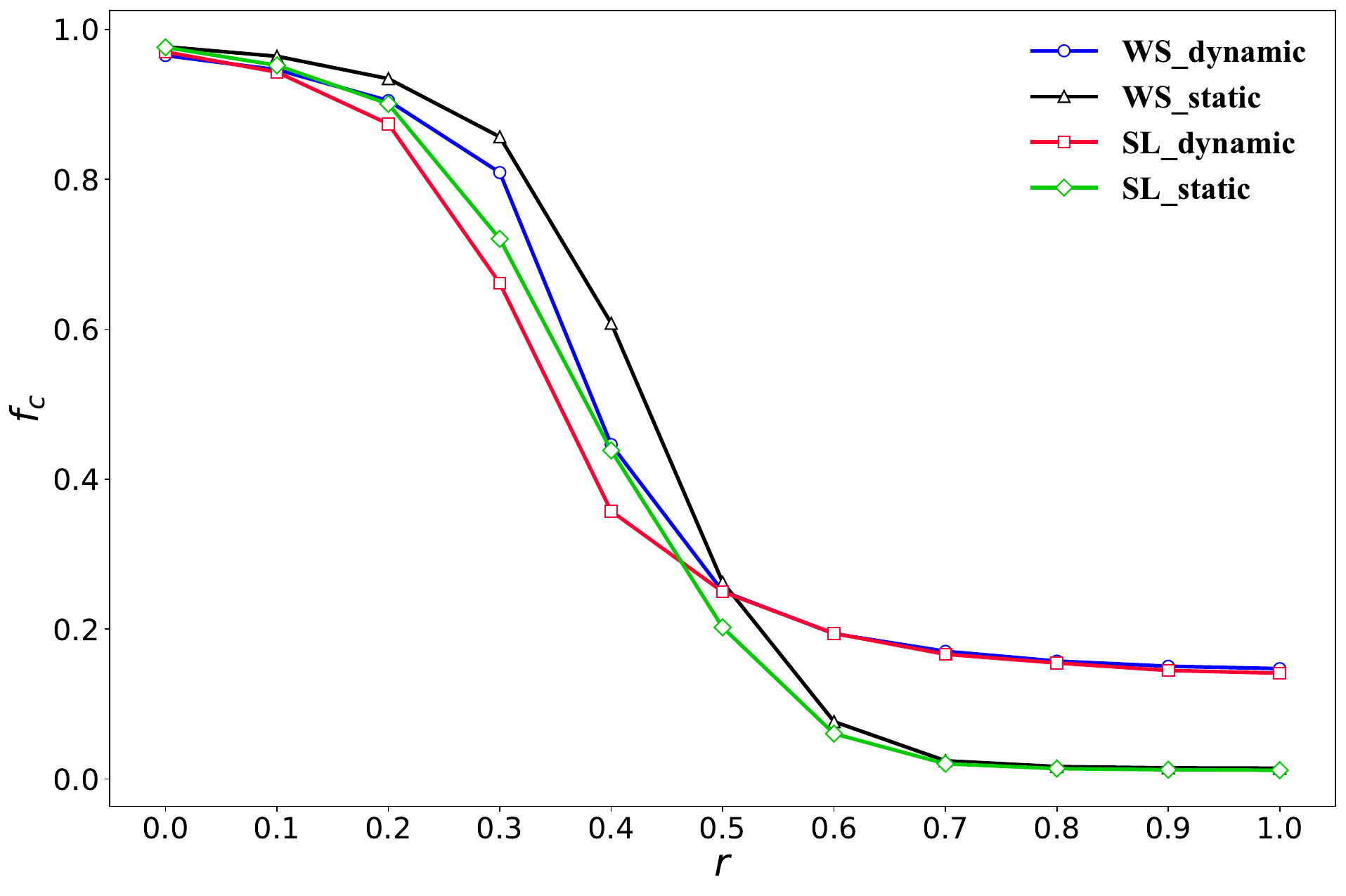}
\caption{\textbf{Dependence of the stationary cooperation fraction $f_c$ on the payoff parameter $r$ in WS and SL networks with and without dynamic participation.} WS\_dynamic and SL\_dynamic denote systems with the dynamic participation mechanism, whereas WS\_static and SL\_static represent the corresponding systems in which all enterprises remain continuously active. The cooperation fraction decreases with increasing $r$ in all four cases, but dynamic participation helps sustain cooperation under less favorable payoff conditions, preventing the complete collapse observed in the continuously active systems.}
\label{active_compare}
\end{figure}
\end{center}

To investigate how dynamic participation affects the evolution of cooperation, we compare the dependence of the cooperation fraction $f_c$ on the payoff parameter $r$ in WS and SL networks with and without the dynamic participation mechanism. The results are shown in Fig.~\ref{active_compare}. Here, WS\_dynamic and SL\_dynamic denote systems with dynamic participation, whereas WS\_static and SL\_static denote the corresponding baseline systems in which all enterprises remain active throughout the evolutionary process.

As demonstrated in Fig.~\ref{active_compare}, the cooperation fraction decreases with increasing $r$ in all four cases. This is consistent with the payoff structure of the snowdrift game, in which increasing $r$ increases the payoff advantage associated with defection while reducing the payoff received by a cooperator interacting with a defector. In addition, under the parameter settings considered here, cooperation on the WS network remains consistently higher than or comparable to that on the SL network. This result indicates that network topology plays an important role in shaping cooperative evolution, with the heterogeneous local connectivity introduced by the small-world structure providing a different interaction environment from that of the regular lattice. More importantly, the effect of dynamic participation depends strongly on $r$. At relatively small $r$, systems without dynamic participation exhibit slightly higher levels of cooperation. As $r$ increases, however, this relationship is reversed. In the absence of dynamic participation, $f_c$ eventually approaches zero at sufficiently large $r$. By contrast, when enterprises are allowed to switch dynamically between active and dormant states according to their payoffs and aspirations, the decline in cooperation is substantially mitigated, and a nonzero level of cooperation, approximately $0.15$ in our simulations, can still be maintained.

This difference arises because dynamic participation continuously reshapes the set of enterprises involved in resource-sharing interactions and therefore modifies the effective local interaction environment experienced by active enterprises. Consequently, cooperative evolution is no longer governed solely by strategy updating on a fixed set of interacting nodes. Although dynamic participation does not necessarily enhance cooperation when resource-sharing conditions are favorable, it becomes increasingly beneficial as the payoff environment becomes less favorable to cooperation. In particular, it prevents the complete collapse of cooperation observed in the continuously active baseline systems. Similar qualitative behavior is observed on both the WS and SL networks, suggesting that the cooperation-preserving effect of dynamic participation under unfavorable payoff conditions is not restricted to a single network topology. These results further demonstrate that incorporating endogenous participation dynamics into enterprise resource-sharing games provides a possible mechanism for sustaining cooperation even when the underlying payoff structure strongly favors defection.

\begin{center}
\begin{figure}[htbp]
\centering
\includegraphics[scale=0.26]{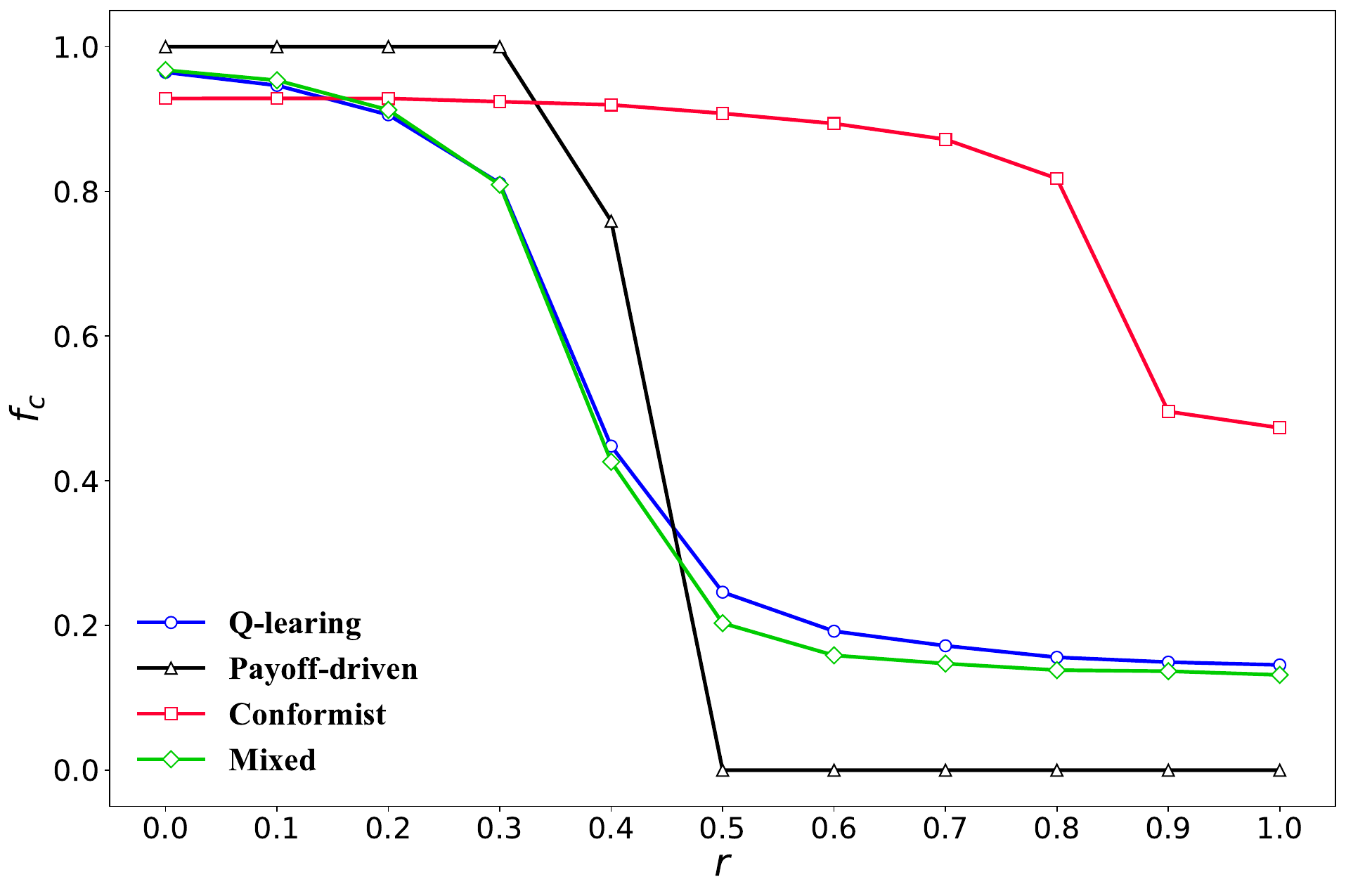}
\caption{\textbf{Comparison of the stationary cooperation fraction under different decision-making modes.} The cooperation fraction $f_c$ is plotted against the payoff parameter $r$ for the Q-learning model, a fixed 50:50 mixture of payoff-driven and conformist enterprises (Mixed), and two homogeneous baselines in which all enterprises adopt payoff-driven or conformist strategy updating. The Q-learning and mixed models exhibit similar cooperation patterns, whereas the homogeneous baselines show markedly different responses to increasing $r$, highlighting the importance of decision-making mode composition.}
\label{Q_compare}
\end{figure}
\end{center}

Next, to assess the contribution of Q-learning to adaptive decision-making, we compare the original model with three alternative settings, including a fixed 50:50 mixture of payoff-driven and conformist enterprises, a homogeneous payoff-driven population, and a homogeneous conformist population. The results are illustrated in Fig.~\ref{Q_compare}. In the mixed model, each enterprise is assigned one of the two decision-making modes with equal probability at initialization and retains that mode throughout the simulation. In the two homogeneous baselines, all enterprises adopt the same decision-making mode. All other model settings are kept unchanged, and the stationary cooperation fraction $f_c$ is examined as a function of the payoff parameter $r$.

As shown in Fig.~\ref{Q_compare}, the Q-learning and mixed models exhibit similar cooperation patterns over the examined parameter range. Both maintain high cooperation for relatively small values of $r$, followed by a pronounced decline as $r$ increases, and the Q-learning model consistently exhibits a slightly higher cooperation fraction than the fixed mixture, although the difference is modest. The homogeneous baselines further demonstrate the importance of the decision-making mode. Under payoff-driven updating, cooperation collapses when $r\geq 0.5$. In contrast, conformist updating sustains a relatively high cooperation fraction over a broader range of $r$, although cooperation also declines at sufficiently large values of $r$. These results indicate that the macroscopic outcome depends strongly on the mode of strategy updating and that the coexistence of different modes can produce cooperation patterns that differ substantially from those observed in homogeneous populations.

\subsection{Robustness of cooperation to network size}

\begin{center}
\begin{figure}[htbp]
\centering
\subfigure[WS]{
\includegraphics[scale=0.25]{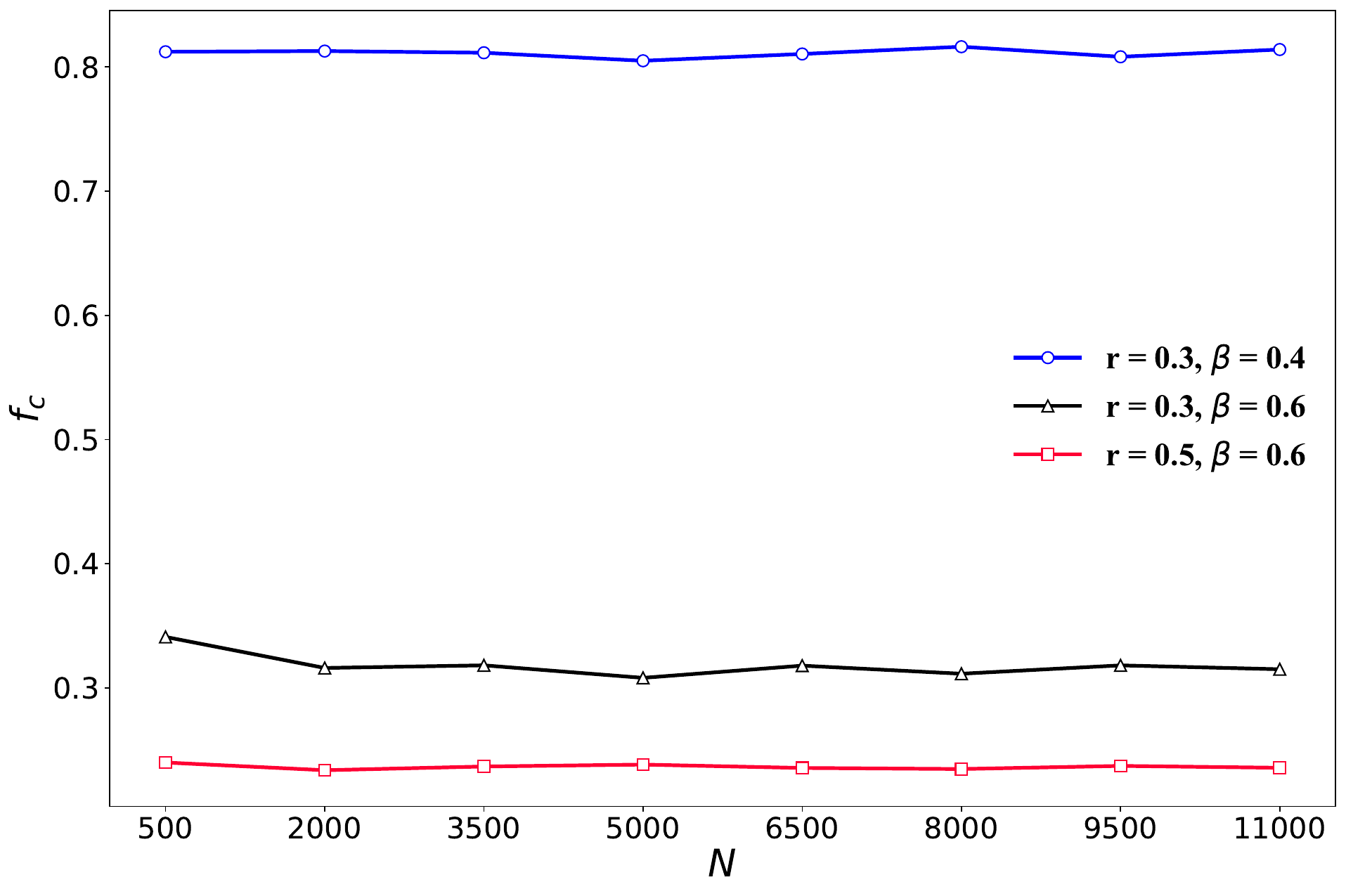}
\label{N_WS}}
\subfigure[SL]{
\includegraphics[scale=0.25]{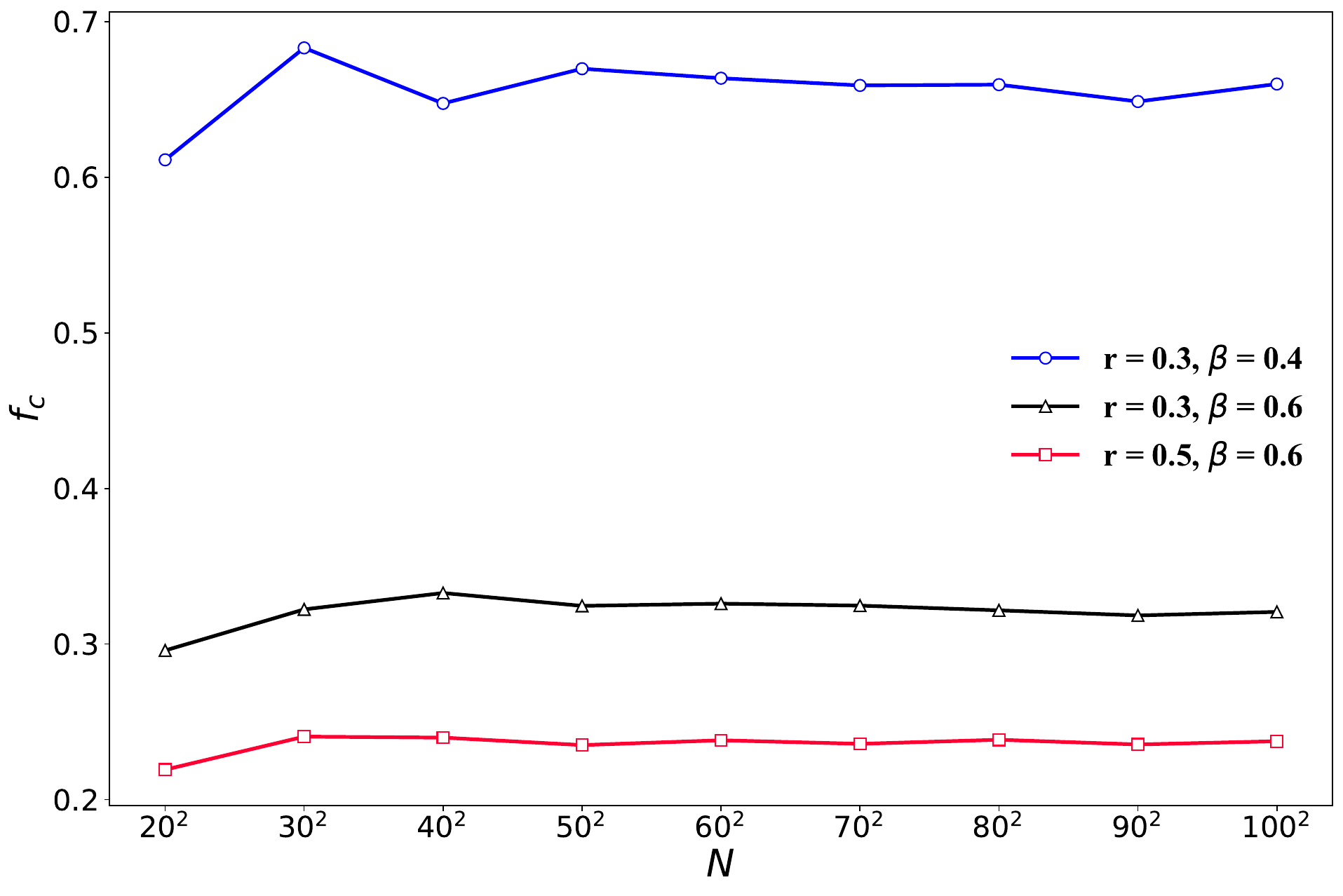}
\label{N_SL}}
\caption{\textbf{Dependence of the stationary cooperation fraction $f_c$ on network size in WS and SL networks.} The three curves correspond to $(r,\beta)=(0.3,0.4)$, $(0.3,0.6)$, and $(0.5,0.6)$. The cooperation fraction exhibits only minor fluctuations with increasing network size in both network structures, which demonstrates the robustness of the cooperation with respect to network size.}
\label{N_fc}
\end{figure}
\end{center}

In real enterprise resource-sharing networks, the system size can be much smaller or much larger than the fixed value of $N=5000$ utilized in previous simulations. Therefore, we investigate the stationary cooperation fraction $f_c$ across different network sizes in both WS and SL networks to examine whether the observed cooperation patterns depend on network size. For WS networks, the network size varies from $N=500$ to $N=11000$. For SL networks, we consider linear lattice sizes from $L=20$ to $L=100$, corresponding to $N=L^2$ nodes. The results are presented in Fig.~\ref{N_fc}, where three payoff parameter combinations are considered, including $(r,\beta)=(0.3,0.4)$, $(0.3,0.6)$, and $(0.5,0.6)$.

From Fig.~\ref{N_fc}, we can see that the cooperation fraction remains relatively stable as the network size increases in both network structures. In WS networks, the three curves exhibit only minor fluctuations over the examined range of $N$, without a clear systematic increase or decrease in $f_c$. A similar pattern is observed in SL networks. Although the cooperation fraction varies somewhat at small lattice sizes, it becomes relatively stable as the system size increases. These results suggest that the observed cooperation levels are not strongly dependent on the specific network sizes considered here.

The three parameter combinations produce distinct cooperation levels in both network structures, indicating that the payoff parameters have a more pronounced influence on the magnitude of $f_c$ than network size within the tested ranges. Moreover, the relative ordering of the cooperation levels associated with the three parameter combinations remains unchanged as the network size varies. By comparing the results, we can see that smaller values of $r$ and $\beta$ are more conducive to the evolution of cooperative behavior, which is consistent with the findings in Fig. \ref{fc_beta_r_WS}.

\section{Conclusion and outlook} \label{Conclusion}

In this paper, we propose an evolutionary game model that integrates dynamic participation, adaptive decision-making, and resource-sharing strategies to investigate the evolution of cooperation in enterprise resource-sharing networks. Each enterprise is characterized by three interrelated attributes, including participation state, resource-sharing strategy, and decision-making mode. Enterprises dynamically switch between active and dormant states according to the discrepancy between their realized payoffs and aspiration levels, adaptively select between payoff-driven and conformist decision-making modes through Q-learning, and update their resource-sharing strategies according to the corresponding behavioral rules. Through extensive simulations on WS and SL networks, we systematically investigate how payoff conditions, aspiration, reinforcement-learning parameters, dynamic participation, Q-learning mechanism, and network size shape enterprise participation and cooperation.

The results reveal some main findings. First, the payoff parameters $\beta$ and $r$ jointly affect both enterprise participation and cooperation, indicating a strong coupling between participation dynamics and strategy evolution. Besides, the aspiration level $A$ has a pronounced non-monotonic effect: intermediate aspiration levels are more favorable for maintaining high levels of participation and cooperation, whereas both excessively low and high aspirations suppress them. Spatial simulations further show that intermediate aspiration levels facilitate the formation and persistence of larger cooperative clusters. Moreover, the stationary active and cooperation fractions vary only slightly over a broad range of the Q-learning parameters $\epsilon$ and $\alpha$, indicating that the qualitative evolutionary outcomes are relatively insensitive to these parameter choices. Furthermore, the effect of dynamic participation depends strongly on the payoff environment. When $r$ is small, continuously active systems can exhibit higher cooperation levels; however, as $r$ increases and the payoff environment becomes less favorable to cooperation, dynamic participation substantially mitigates the decline in cooperation and prevents its complete collapse. Similar qualitative behavior is observed on both WS and SL networks. The ablation study further clarifies the role of Q-learning in the proposed framework. Although Q-learning yields a slightly higher stationary cooperation fraction than the fixed mixture under the examined conditions, its additional contribution to cooperation is relatively modest. The homogeneous baselines further demonstrate that the choice of strategy-updating mode can substantially influence cooperation. Additionally, simulations across different network sizes reveal that the stationary cooperation fraction remains relatively stable in both WS and SL networks within the examined size ranges, which supports the robustness of the observed cooperation patterns to network size. Overall, these findings highlight the role of endogenous participation in shaping cooperation through its interaction with strategic behavior and adaptive decision-making. Participation is not simply a prerequisite for resource-sharing interactions, but an evolving behavioral dimension that determines which enterprises are currently engaged in the game. Changes in participation consequently alter the effective interaction environment, while the strategies retained by dormant enterprises influence their behavior when they become active again. In this way, participation dynamics are coupled with cooperation strategy evolution and adaptive selection between decision-making modes, jointly shaping the emergence and persistence of cooperation. The presented results therefore suggest that cooperation in resource-sharing networks can be understood more comprehensively by considering the interplay between participation, strategy updating, and behavioral adaptation, rather than examining these processes independently.

Although the proposed model can describe the adaptive evolution across participation, decision-making, and cooperation in resource-sharing networks, several extensions deserve further investigation. On the one hand, the present model assumes a static underlying network in which potential resource-sharing relationships remain unchanged throughout the evolutionary process. In real enterprise systems, however, enterprises may establish new relationships, terminate existing ones, or switch partners according to their previous interaction outcomes. Future work could therefore incorporate link formation, link removal, and partner selection to study the coevolution of network topology, participation, and cooperation. Such extensions would allow the interaction structure itself to respond to participation and strategic outcomes, providing a more comprehensive perspective on the coevolution of network topology and behavioral dynamics. On the other hand, enterprises in the model follow relatively homogeneous payoff structures, whereas real enterprises differ substantially in size, resource endowments, risk preferences, resource demands, and aspiration levels. Introducing such heterogeneity, together with more realistic payoff structures, may further improve the model's ability to represent real-world resource-sharing systems. Moreover, the payoff-driven and conformist updating rules considered in this study represent only two possible mechanisms of behavioral adaptation. Future research could incorporate alternative strategy-updating rules to capture a broader range of decision-making processes and investigate how different behavioral mechanisms interact with endogenous participation to shape cooperation in resource-sharing networks.

\section*{Acknowledgments}

This work was supported in part by the Natural Science Foundation of Chongqing under Grant No. CSTB2025YITP-QCRCX0007, in part by the Fundamental Research Funds for the Central Universities under Grant No. SWU-KT26010, and in part by the National Research, Development and Innovation Office (NKFIH) under Grant No. K142948.

\section*{Data availability}
The data that support the findings of this study are available from the corresponding author upon reasonable request.




\section*{References}
\bibliographystyle{unsrt}
\bibliography{aipsamp.bib}


\end{document}